# First Results from an Event Synchronized - High Repetition Thomson Scattering System at Wendelstein 7-X

H. Damm,[a,1] E. Pasch,[a] A. Dinklage,[a] J. Baldzuhn,[a] S.A. Bozhenkov,[a] K.J. Brunner,[a] F. Effenberg,[b] G. Fuchert,[a] J. Geiger,[a] J.H. Harris,[c] J. Knauer,[a] P. Kornejew,[a] T. Kremeyer,[b] M. Krychowiak,[a] J. Schilling,[a] O. Schmitz,[b] E.R. Scott,[a] V. Winters[b] and the Wendelstein 7-X Team[2]

[a]*Max-Planck-Institut für Plasmaphysik,
D-17491 Greifswald, Germany*

[b]*University of Wisconsin,
Madison, Wisconsin 53706, USA*

[c]*Oak Ridge National Laboratory,
Oak Ridge, Tennessee 37831, USA*

E-mail: hannes.damm@ipp.mpg.de

Abstract: The Wendelstein 7-X Thomson scattering diagnostic was upgraded to transiently achieve kilohertz sampling rates combined with adjustable measuring times. The existing Nd:YAG lasers are employed to repetitively emit "bursts", i.e. multiple laser pulses in a short time interval. Appropriately timing bursts in the three available lasers, up to twelve evenly spaced consecutive measurements per burst are possible. The pulse-to-pulse increment within a burst can be tuned from $2\,\mu$s to $33.\overline{3}\,$ms (500 kHz - 30 Hz). Additionally, an event trigger system was developed to synchronize the burst Thomson scattering measurements to plasma events.
Exemplary, a case of fast electron density and temperature evolution after cryogenic $H_2$ pellet injection is presented in order to demonstrate the capabilities of the method.

Keywords: Plasma diagnostics - interferometry, spectroscopy and imaging; Trigger concepts and systems; Detector alignment and calibration methods, Lasers

---

[1]Corresponding author.
[2]Full author list: T. Klinger et al., *"Overview of first Wendelstein 7-X high-performance operation"*, doi.org/10.1088/1741-4326/ab03a7, Nucl. Fusion (2019).

# Contents

## 1 Introduction

The magnetic confinement fusion experiment Wendelstein 7-X [1] (W7-X) is one of the largest stellarator experiments [2] world wide. W7-X is a helical-axis advanced stellarator (HELIAS) [3]. It was optimized with regard to engineering and physics objectives [4–6] to demonstrate the viability of the stellarator concept as a candidate for future fusion power plants. Major physics results during the progressing project were summarized in individual publications for the machine commissioning campaign [7–10] in 2015 and the first physics campaigns [11, 12] in 2017 and 2018. While W7-X aims at quasi steady-state operation (i.e. up to 30 minutes), transients are crucial both for the development and the control of fusion relevant plasma scenarios. One example is the fueling of the plasma core [14, 15] via the injection of cryogenic $H_2$ pellets [13], which will be employed as working example throughout this paper.

Electrons play a central role for the fueling since the energy is transferred through electron collisions in the pellet deposition process. Thus, the fast evolution of electron density ($n_e$) and temperature ($T_e$) is of high interest. Incoherent Thomson Scattering (TS) [16] provides a well-established approach to measure the spatio-temporal evolution of both. Requiring high laser power, TS of plasma electrons is usually conducted with pulsed laser systems. This stroboscopic measurement results in a regular time comb of data points. Short unpredictable events would only be captured by coincidence and cannot be studied systematically, because timescales faster than the inverse repetition frequency of the laser system (in the case of W7-X: $33.\bar{3}$ ms for 30 Hz) are not resolvable. This problem is visualized in Figure 1, comparing standard W7-X TS measurement times (black lines) in a), to the desired measurement times (dashed gray lines) in b) and c).

This paper discusses the implementation of a technique to deliver highly sampled $n_e$ and $T_e$ measurements (in a short but well-defined time period) demonstrated on experiments in W7-X. The method relies on a fast sequence of consecutive laser pulses [17] together with the synchronization of the laser pulses to the event of interest [18, 19]. The high repetition operation "burst" mode will be described in Section 2 after a brief introduction to the W7-X TS diagnostic; followed by the event synchronization system in Section 3. As an example application, transient $T_e$ and $n_e$ burst mode profile measurements of the ablation and deposition of a cryogenic $H_2$ pellet are briefly discussed in Section 4; and finally Section 5 summarizes the this work, accompanied by a short outlook.

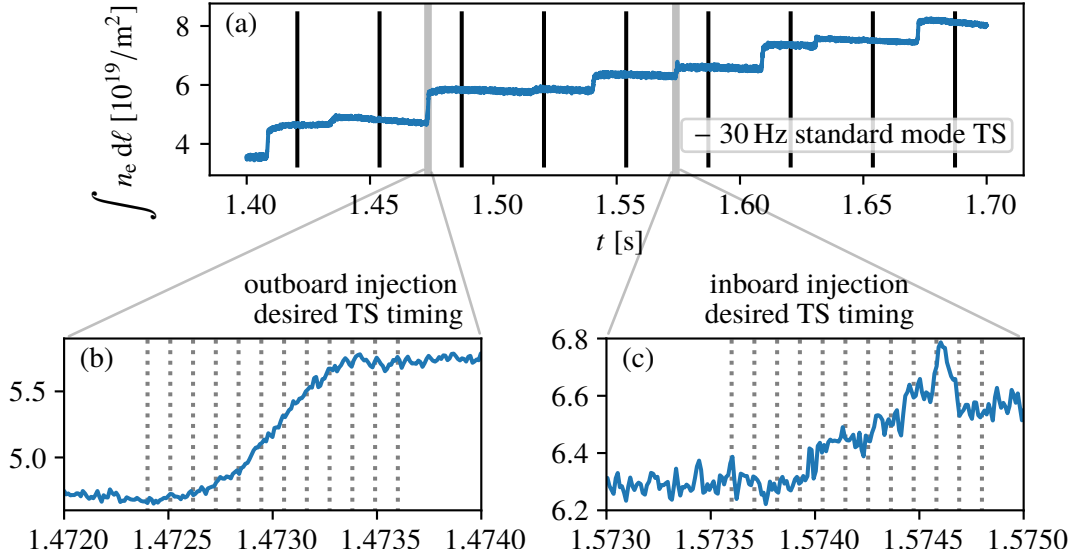

**Figure 1**: a) Stepwise increase of the line-integrated density due to consecutive pellet injetion in W7-X program 20180918.044. The black vertical lines mark the timings of the measurements taken in the normal TS mode. b) and c) show a zoomed view to different pellet injections; the dashed grey lines indicate the desired timings for resolving the plasma event by Thomson scattering measurements with a $\Delta t = 100\,\mu s$ time-comb.

## 2 Increasing the Laser Repetition Rate

The TS diagnostic of W7-X (detailed setup and specifications in [20, 21]) employs three Q-switched Nd:YAG (neodymium-doped yttrium aluminum garnet) high power lasers "SplitLight2500" from the manufacturer "InnoLas Laser GmbH". The nominal repetition frequency of each laser is 10 Hz and the maximum energy per pulse 2.50 J. (The lasers are usually not run at full power in the standard mode, because the signal from the plasma center would saturate at higher laser energies for typical plasma densities. Pulse energies of 1.5 J have proven to be a good compromise avoiding saturation of the central signal but yet maintaining acceptable signals from the plasma edge.) The individual lasers can be triggered either alternatingly to provide higher total repetition frequencies or simultaneously to increase the accumulated pulse energy. Two optics collect the laser photons scattered by the plasma electrons into optical fibers. Along the laser beam path 42 scattering volumes are aligned approximately in radial direction of the plasma torus and thus provide a full profile. The distance between the volume centers ranges from 0.7 cm to 7.8 cm in real space coordinates. The analogue digital converter "SP Devices ADQ-14" employed in the W7-X TS system features a constantly recording ring buffer. A 2 $\mu s$ cycle duration was chosen for the buffer which defines the minimum required delay between two consecutive TS measurements. In the post-processing, the spectral distribution of the scattered light is reconstructed from the measured signal. The amplitude of the spectrum scales to the electron density (and the incident laser power) whereas the spectral blue shift and shape are employed to determine the electron temperature [16].

Neodymium, which is employed in the W7-X TS lasers, provides an energetic four-level electron orbit system. The life time of its meta-stable $^4F_{3/2}$ level is $\tau = 230\,\mu s$ [22]. 200 μs flash lamp (FL) flashes are used in the SplitLight2500 lasers for optical pumping. The laser light is released from the Nd:YAG rod after applying a high voltage to the Pockels cell (PC) acting as Q-switch. The timing scheme of this process is shown in Figure 2a).

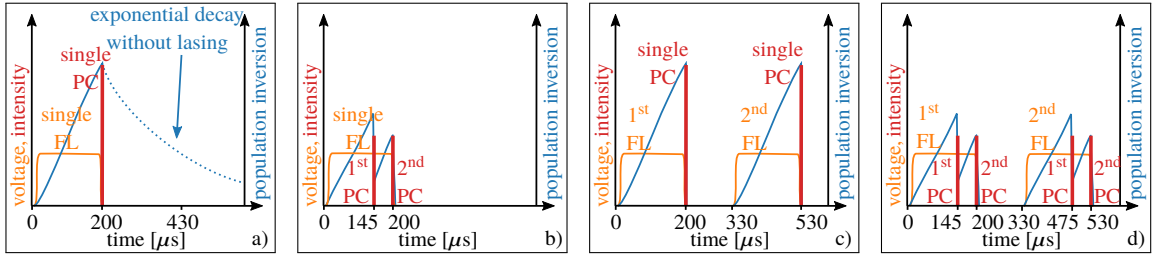

**Figure 2**: Timing scheme of the internal laser components (FL = flash lamp, PC = Pockels cell) for a) standard operation, b) double PC operation - The energy stored in the population inversion introduced by the FL flash is split into the two pulses extracted by two PC openings, resulting in halved energy per laser pulse., c) double FL operation - The laser base-repetition frequency is reduced from 10 Hz to 5 Hz, but the single FL cycle is replaced by two cycles. and d) combined double PC & double FL operation.

Two different ways to increase the repetition rate of the lasers were employed. The first option was to split the energy from a single FL flash (stored in the population inversion of the electrons in the lasing medium) into multiple laser pulses via multiple PC openings as depicted in Figure 2b). The amount of pulses generated by this method is only limited by the PC-response time. Nevertheless, increasing the number of pulses comes at the cost of lower individual pulse energies. This is an expensive trade off for additional measurements, because the intensity of the scattered light and thus the TS signal to noise ratio is proportional to the incident laser power. For this reason, in the diagnostic application no more than two PC openings per FL cycle were used. For the PC double pulse operation the lasers are run at full power which results in 1.25 J per pulse. To reach this energy, the maximum delay between double PC pulses within one FL cycle is approximately 166 μs limited by the population inversion lifetime and the flash lamp discharge duration.

The second option used to generate two consecutive laser pulses within a short time interval was the reduction of the base repetition frequency of the lasers by a factor of two, replacing the single FL cycle by two cycles as shown in Figure 2c). The average number of FL flashes and thus the heat load in the laser remains constant. The advantage of this method compared to the former one is the maintained pulse energy. Similar to the double PC operation, the time between two pulses is limited by the the duration of the FL discharges and the population inversion lifetime, but in the case of double FL operation, they define a lower limit. Too low temporal spacing would result in spontaneous laser emission during the second FL cycle. This phenomenon is not yet understood. A lowest possible time distance for double FL pulses was experimentally determined to be 330 μs. An upper limit of 5 ms for the double FL increment is set by the laser software rather than the physics and can be extended by the manufacturer on request. Higher number of multi FL pulses at correspondingly lower base repetition frequencies are currently under investigation. The key limiting factors (in the order of precedence) are: i) the thermal load which introduces stress to the rods that could cause damage, ii) the change of the thermal lens which could introduce hot spots in the beam profile and iii) the reduced voltage in the capacitor banks of the FL which cannot be reloaded quick enough and thus lead to changing energies per laser pulse.

By combining the two introduced methods as shown in Figure 2d), a pulse train of four pulses emitted from one laser is possible. With the three available W7-X TS lasers, this pattern can be extended to a train of 12 pulses on a 5 Hz basis. The upper and lower restrictions by the FL pulse duration and population inversion allow for uniform increments of $55 - 166\,\mu$s for the 12 pulse burst mode. It thus can provide burst TS data in the $6.\overline{6} - 18.\overline{18}$ kHz range for a duration of $0.605 - 1.826$ ms (i.e. the total duration of the selected burst) every 200 ms, corresponding to the 5 Hz laser base repetition frequency. Many other patterns of finite number consecutive laser pulses with wider or narrower temporal spacing are possible combining the double FL and the double PC operation mode, as well as non-burst modes and/or non-uniform timings. A large subset of the possible operation modes was experimentally qualified for safe operation with the W7-X TS setup and is summarized in Table 1. Bursts of $3 - 12$ pulses with temporal spacing tune-able from $2\,\mu$s to $33.\overline{3}$ ms for different experimental requirements on a 5 Hz or 10 Hz basis are covered.

**Table 1**: W7-X Thomson scattering "burst" operation modes. The 30 Hz standard operation mode is covered as well (3 pulses alternated with $33.\overline{3}$ ms delay on a 10 Hz basis).

| laser operation | standard | double PC | double FL | double PC & FL |
| --- | --- | --- | --- | --- |
| burst/event duration | $4\,\mu$s – $66.\overline{6}$ ms | $10 - 750\,\mu$s | $0.55 - 25$ ms | $0.605 - 1.826$ ms |
| pulses per burst | 3 | 6 | 6 | 12 |
| temporal spacing | $2\,\mu$s – $33.\overline{3}$ ms | $2 - 150\,\mu$s | $110\,\mu$s – 5 ms | $55 - 166\,\mu$s |
| event sync. | in preparation | in preparation | yes | yes |
| energy per pulse | 1.5 J (up to 2.5 J) | 1.25 J | 1.5 J (up to 2.5 J) | 1.25 J |
| base rep. frequency | 10 Hz | 10 Hz | 5 Hz | 5 Hz |

The energy monitor of the standard mode TS system only supports $\approx 10$ Hz acquisition frequencies. Therefore, a 18 V biased Si-PIN diode of type "PDSIU500FC1D-W-0" (active area of $500\,\mu$m$^2$) from "PL-LD Inc." supporting readout frequencies of up to $\approx 600$ MHz was added to the setup. It is employed for the energy correction of the density signal if the TS diagnostic is run in the burst mode.

## 3 Realization of the Event Synchronization

Employing the high repetition modes described in Section 2 frequently occurring short events like ECCD crashes [23] and precisely predictable short events (neutral beam [24] and electron cyclotron resonance [25] heating modulation and impurity laser blow-off [26]) have been studied and will be published elsewhere. To study randomly occurring events or events suffering from a large jitter (H$_2$ pellet injection: up to 30 ms [13]), the individual burst mode pulse trains must be synchronized to the event.

Although the base repetition frequency of the pulsed lasers in principle is fixed to 5 or 10 Hz, the slow nature of the thermal processes in the laser rods was exploited to make event synchronized lasing possible. The characteristic time for thermal changes in the SplitLight2500 laser rods $\tau_{\text{thermal}} = r_{\text{rod}}^2 \rho c / 4\kappa$ (Nd:YAG rod: radius $r_{\text{rod}} = 6.5$ - $12$ mm, density $\rho \approx 4.56 \times 10^6$ g/m$^3$, thermal conductivity $\kappa \approx 12.9$ W/mK, specific heat capacity $c \approx 0.59$ Ws/gK [27]) lies in the range of seconds, as can be derived from the General Heat Equation [28]. This allows for non-uniformly distributed FL flashes below this thermal timescale, if the average base repetition frequency is maintained. For the W7-X TS lasers it was experimentally verified, that consecutive bursts can be triggered with time displacements alternating back and forth between 30 ms and 370 ms as compared to the constant 200 ms displacements for uniform 5 Hz operation. This finding allowed for the design of the W7-X TS event trigger logic which, together with a pulse generator:

1. provides a 5 Hz standard trigger,
2. blocks event triggers too close ($\leq 30$ ms) to the standard trigger,
3. accepts event triggers in the safe interval and skips the next standard trigger after an event,
4. blocks multiple events in one acceptance interval (maintain average repetition frequency).

These constraints are visualized in Figure 3a) and the simplified technical implementation is shown in Figure 3b).

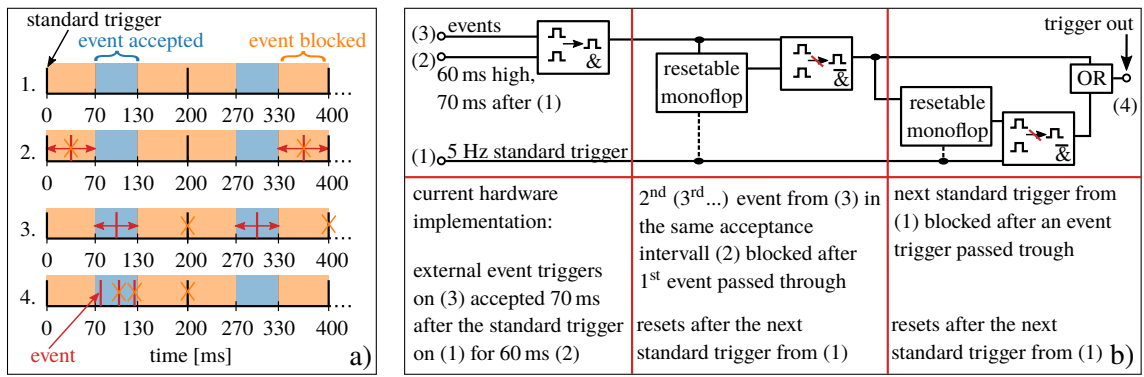

**Figure 3**: a) The four requirements to the W7-X TS event trigger logic: 1. Providing a 5 Hz standard trigger, 2. blocking event triggers too close to the standard trigger (orange background regions), but 3. accepting event triggers in the safe interval (blue background regions) and skipping the next standard trigger after an event, as well as 4. blocking multiple events in one acceptance interval (to maintain the average repetition frequency).
b) The simplified technical implementation of a) employing a hardware logic circuit. The maximum displacements or in other words the extends of the event trigger acceptance interval can be adjusted via the signal applied to channel (2), because the AND only allows an event from (3) to pass, if both inputs are at "high" level. The first monoflop/NAND-combination will block further event triggers until the next standard trigger from (1). The second set will block the next standard trigger after an event passed trough. Together, they maintain the average base repetition frequency of the laser.

The hardware setup was employed for a proof of principle in the last operational campaign. It allowed for displacements of the bursts alternating between 70 ms and 330 ms, which stayed safely within the verified 30 ms and 370 ms displacements. If no event occurred, the bursts were triggered automatically with 200 ms displacements (5 Hz) by it. A software implementation based on the Field

Programmable Gate Array (FPGA) of the W7-X "Trigger-Time-Event-System" [29] was developed recently to replace the hardware. It will allow for 10 Hz operation event synchronization (100 ms displacements on average) in addition to the current 5 Hz mode in the future and for switching between burst modes and the standard mode remotely.

Currently, the ablation radiation of TESPEL [30] and $H_2$ pellets can be used as trigger signals which are transmitted to the setup via an optical fiber link. The cable transmission time and the internal electronics delay add up to about 8 μs. This integral delay is much shorter than the laser's FL pumping time of about 200 μs and therefore is negligible.

## 4 A Possible Application: Cryogenic $H_2$ Pellet Injection

The motivation for the developing the event synchronized high repetition mode for the W7-X TS was to understand the ablation and deposition physics mechanisms of cryogenic $H_2$ fueling pellets [31]. The average duration of the pellet ablation is known from the ≈1 ms extent of the $H_\alpha$ ablation emission originating from excited neutrals of the ablated pellet material. Since the blower-gun injector [32] cannot provide precise control over the pellet velocity, the injection timing suffers from a jitter which is an order of magnitude larger than the ablation duration. This made an event synchronization necessary. To study the ablation and fast deposition phase of the pellet by a burst measurement, the 12 pulse burst mode with 100 μs pulse-to-pulse increment covering a 1.1 ms interval was employed for most of the pellet injection experiments.

The ablation emission detected by a filterscope [33] and the line-integrated density signal from the interferometry diagnostic [34] were employed to verify the match of the burst TS measurement to an outboard side pellet injection into W7-X program 20180918.046 as shown in Figure 4 (c.f. desired TS timings shown in Figure 1).

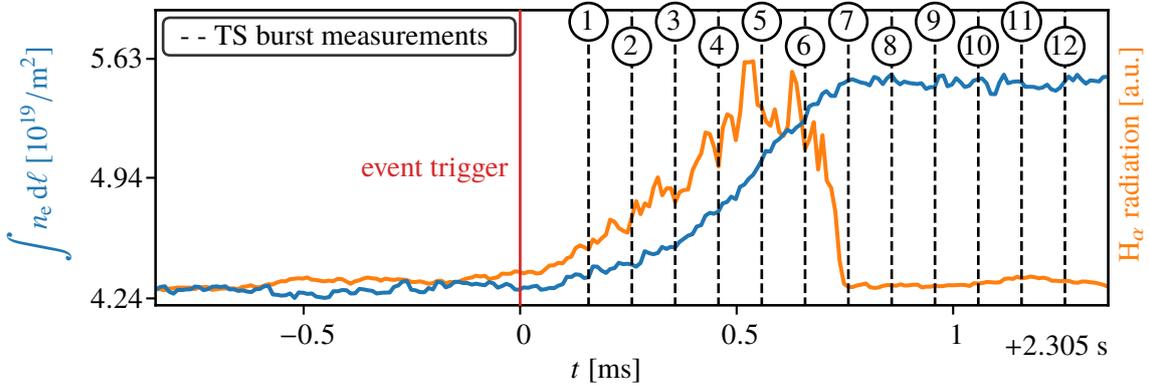

**Figure 4**: Twelve time points (black dashed vertical lines) relating to a burst of event triggered TS data taken during the W7-X program 20180918.046 after an event trigger (red vertical solid line) at ≈2.305 s. The comparison to the pellet $H_\alpha$ ablation signal (orange) and the line-integrated density increase (blue) introduced by the pellet confirm a successful match of event and measurement.

As can be deduced from the figure, the TS measurement timing perfectly matches the plasma event and even extends beyond the line-integrated density step. This is intentional, because although the total plasma particle number does not increase anymore after complete ionization of the pellet material, redistribution effects e.g. due to particle transport are still observed in the radially resolved TS burst mode data shown in Figure 5.

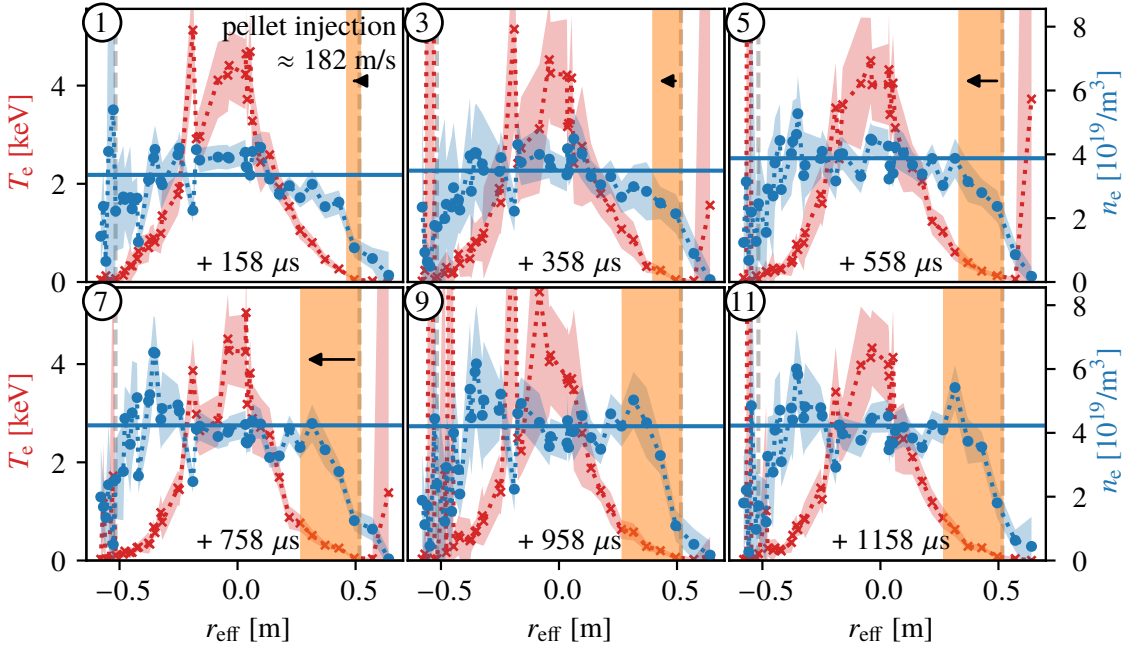

**Figure 5**: The figure shows every other TS measurement (electron density blue dots, electron temperature red crosses) of the burst triggered to a H$_2$ pellet injected from the outboard side of the torus in W7-X program 20180918.046 at $\approx 2.305$ s as shown in Figure 4. A label in each sub-figure shows the timing of the measurement relative to the event trigger. The outboard and inboard side scattering volumes are mapped onto positive and negative effective radii respectively. The error bands relate to statistical $2\sigma$ errors and dotted lines were added to guide the eye between the measurement points. The location of the last closed flux surface of the plasma is indicated by the dashed gray line and the horizontal solid blue line relates to the line averaged interferometry density (line-integrated signal normalized to the integration path). The injection direction and momentary penetration depth of the pellet are indicated by the black arrow (vanishes after complete ablation of the pellet); the orange area covers the region of the plasma in which pellet material was ablated. From the plasma edge inwards up until $r_{\text{eff}} = 0.3$ m the density increases locally, when comparing density profile no. 1 with no. 7, 9 and 11. This indicates the local deposition of particles after the pellet ablation. The profile shape changes from a flat/convex shape to a slightly hollow one. The electron temperature drops locally, i.e. comparing temperature profile 1 and 7, aligned with the local density increase.

For the first time at W7-X these measurements resolved the temporal change of the density profile shape from flat/convex to hollow (symmetric density "wings" at $r_{\text{eff}} \approx \pm 0.4$ m) after a cryogenic H$_2$ pellet injection on a 1 ms timescale. A temperature decrease can be identified, e.g. by comparing measurements 1 and 7, which is aligned with the local density increase.

The kinetic energy stored in the plasma electrons $W_e = \int p_e dV \approx 300$ kJ, with $p_e$ being the electron pressure, stayed constant within the uncertainties of the measurement on the 1 ms timescale after injection of the pellet. The energy input from the electron cyclotron resonance heating in the same time interval (3.2 kJ) was smaller than the uncertainties of the $W_e$ measurement and therefore negligible. Moreover, the analyzed timescale is much shorter than the expected transport timescales for the given type of plasma. Accordingly, the analyzed process must be adiabatic, as supported by the data, which in turn confirmed the validity of the measurement.

Due to a misalignment of the TS laser beams with respect to the calibration measurement in the last campaign, absolute calibration factors for the density data are not available. Instead, the TS line integral was set to the value gained by the interferometry diagnostic which shares the TS laser line of sight.

## 5 Summary and Outlook

A flexible trigger system and a short-duration high repetition frequency mode was introduced to the Wendelstein 7-X Thomson scattering diagnostic. It can be employed to match individual experimental requirements, in addition to the 30 Hz standard mode. The new mode features short sequences of laser pulse with flexible pulse-to-pulse temporal spacing. It was implemented using specific double Pockels cell and double flash lamp operation modes of commercially available Nd:YAG lasers cycled with 5 or 10 Hz. Additionally, a trigger system was developed to allow for safely firing the TS lasers with a certain variability sustaining the average laser repetition frequency. It is able to match the burst of TS measurements to short transient plasma events. If no event occurs, it triggers the TS lasers uniformly. Both upgrades were introduced without major changes to the W7-X TS diagnostic which now allows for time resolved electron density and temperature profile studies of short-lived predictable and unpredictable plasma events. As an example, a burst measurement of an outboard side cryogenic $H_2$ pellet injection was shown. The successful temporal match of event and measurement was demonstrated and transient electron density and temperature changes introduced by the plasma event were detected.

In-depth analysis of the now accessible physics like pellet penetration and localized deposition studies, particle transport and gradient evolution or inboard-outboard injection profile comparisons is work in progress. Once the trigger timer event system becomes fully available for the post-2020 experimental campaigns, more unpredictable events can easily be added as triggers. Together with the software version of the event trigger logic available in the near future the TS diagnostic can be employed to record almost any plasma event of interest. A multi-sigthline interferometer [35] is planned to be added to W7-X. It will observe the plasma at a different toroidal location than the TS diagnostic. Inverted fast sampled density profiles from this diagnostic can be compared to the the burst mode measurements. The combination of the two diagnostics will be a unique system to study toroidal features of short plasma events.

## Acknowledgments

The authors would like to thank the InnoLas Laser GmbH for the support exploring the capabilities of the laser system beyond the standard operation mode.

This work has been carried out within the framework of the EUROfusion Consortium and has received funding from the Euratom research and training programme 2014-2018 and 2019-2020 under grant agreement number 633053. The views and opinions expressed herein do not necessarily reflect those of the European Commission.

**Note added.** The data analysis and processing codes used for this paper can be found in `git.ipp-hgw.mpg.de/hdamm/papers` - git commit 568080b63c758f88fa64145c72afd2739632ac9c and `git.ipp-hgw.mpg.de/hdamm/ts_archive_2_profiles` - git commit 8079f1aab7f3bcaf70f6e66ba1d502a6b1336785.